\documentclass[conference, 10pt]{IEEEtran}
\IEEEoverridecommandlockouts

\usepackage{cite}
\usepackage{amsmath,amssymb,amsfonts}
\usepackage{algorithmic}
\usepackage{graphicx}
\usepackage{textcomp}
\usepackage{xcolor}
\usepackage{epsfig}
\usepackage{mystyle}
\usepackage{LGI}

\def\BibTeX{{\rm B\kern-.05em{\sc i\kern-.025em b}\kern-.08em
    T\kern-.1667em\lower.7ex\hbox{E}\kern-.125emX}}

\begin{document}

\title{Scalable, Secure and Broad-Spectrum Enforcement 
of  Contracts---Without Blockchains\\
{\footnotesize \textsuperscript{*}}
}

\author{\IEEEauthorblockN{1\textsuperscript{st} Naftaly Minsky}
\IEEEauthorblockA{\textit{Computer-Science Department} \\
\textit{Rutgers University}\\
New Brunswick NJ, USA \\
minsky@cs.rutgers.edu }
\and
\IEEEauthorblockN{2\textsuperscript{nd}Chen  Cong}
\IEEEauthorblockA{\textit{Computer-Science Department} \\
  Rutgers University \IEEEauthorrefmark{10}{(Now employed by Google LLC)}\\ 
New Brunswick, NJ, USA \\
chen.cong@cs.rutgers.edu
}
}

\maketitle

\begin{abstract}
  This paper introduces a scalable and secure contract-enforcement mechanism,
  called Cop, which can be applied to a broad range of multi-agent systems including small
  and large systems,  time-critical systems, and
systems-of-systems.
 Cop enforces contracts  (or protocols) via the existing Law-Governed
Interaction (LGI) mechanism,
coupled with a new protective layer that significantly enhances the
dependability and security of such enforcement.

Cop is arguably superior to the currently popular
blockchain-based smart-contract mechanisms, due to its scalability,
interoperability, and the breadth of the spectrum of its domain of  applications.
\end{abstract}

 
\section{Introduction}
 There are many situations where a group  of autonomous  actors---which  may be
 software processes, physical devices, and/or people operating via a platform
 like a smart-phone---need to interact with each other over the Internet.
The members of such a group, who may not trust each other, may be required to
interact subject to a given protocol. Such a protocol
 may represent a contract that binds these actors, or it may
 be necessary for the actors to collaborate effectively on some common goal, or
 to compete safely over the use of some resources.
 We call a group of actors that
 comply with a given protocol P, a \emph{P-community}, or simply a community.

It is sometimes possible to establish a given protocol P over a community by relying
on voluntary compliance with it, by all the members of this community.
 Voluntary compliance can be effective when a community is homogeneous, or when
 it is relatively small and its members trust each other, or when its members
 are well managed.
 But otherwise, for voluntary compliance with a given protocol P to be effective it needs to satisfy
  the following two conditions, as argued in   \cite{min03-7}:
(a) it must be the vested interest of every member of the community to comply
    with P; and (b) a failure to
comply with P, by anybody,  should not cause any serious harm to
anybody else in the given community.
And if any one of these conditions is not satisfied,  then the given protocol may need to be
\emph{enforced}.  The enforcement of protocols is the subject of this paper.
And we call a mechanism used for enforcing protocols
 a \emph{protocol enforcement mechanism} or PEM.

We start in  \secRef{quality} by introducing the set  of  qualities of a PEM
that we consider essential. We continue in  \secRef{art} by reviewing two
existing realizations of PEM: (a) the currently popular   blockchain-based smart-contract  mechanism, and (b) the older law-governed interaction (LGI)
mechanism introduced by the 1st author---evaluating them in terms of these
qualities. We find that each of
these mechanisms fails to satisfy  some of these qualities.

We obtain a  PEM that satisfies all these qualities, by extending LGI into
what we call \emph{Cop}.   Before introducing Cop we outline, in  \secRef{LGI}, the LGI
mechanism itself, and describe its shortcoming---which we
attempt to resolve. The LGI-based
Cop mechanism is introduced in \secRef{Cop}, and discussed further in
subsequent sections. We conclude in \secRef{conclusion}, and  with an  appendix
in \secRef{append}.

\paragraph{A Terminological Comment}
In the computer-science literature the concept of protocol is sometimes referred
to  as a contract, and sometimes as a law.   All three terms are being used
here in the following way: We use the term ``protocol''
 as a general term, when not discussing any particular enforcement
mechanism; we use  the  term ``contract'' when discussing blockchain-based
smart-contracts; and we use the term ``law'' when discussing LGI and Cop.

\section{The Essential Qualities of Protocol-Enforcement Mechanisms}\label{quality}
There is a wide range of types of applications
that may benefit from the enforcement of protocols. Such applications differ along
several dimensions. Some are small, others may be very large. Some are simple
enough to be handled as a single community operating subject to a single
protocol,
others consist of multiple
communities, operating subject to different protocols, which may need to
inter-operate in various ways. Some applications are lax about the speed of
protocol enforcement, others are time-critical.
The actors involved in such applications may include software processes,
people, and
physical devices (i.e., devices that belong to IoT).

 One would like to have a
single protocol-enforcement mechanism (PEM) which is sufficiently
broad-spectrum to support a wide range of potential applications.
We identify below  four qualities
that a PEM should satisfy to be sufficiently broad-spectrum.

\begin{enumerate}
\item \emph{Sufficiently short latency}: By latency we mean, the time it takes for a PEM to resolve a given
transaction\footnote{We use the term ``transaction'' to mean any interactive
operation by one of the actors in a given community.}.
The maximal latency that a given application may require varies widely.
It may, for example, be of the order of a few minutes for many commercial and
financial systems. But it may be of the order of milliseconds or less for 
the so-called \emph{time-critical} applications---such
 as a collection of interacting
physical machines operating in an industrial plant, or the
components of an airborne control system interacting with each other.

\item \emph{Scalability:} By scalability we mean here that the latency is virtually
      independent of the volume and frequency of transactions in a given system.
 Scalability is a challenge to large systems, such as enterprise systems, and
to  infrastructures such as air-traffic control systems. It is also a challenge to
 financial systems such as the one described in \cite{bla18-1}, where
"a reasonable estimate of [its] peak figure may be in the region of several
thousand transactions per second."

\item \emph{Interoperability:} By  interoperability we  mean: the ability of
communities that operate subject to different protocols to interact with each
other. Of course,  such an ability needs to be subject to regulation. That is,
there needs to be a way to control which communities can
interact with each other, and how.

 Interoperability is required, in many situations.  In particular, when
 different  small businesses, each operating under its own protocol, need to interact
with each other.  Moreover, complex systems, such as enterprise systems, cannot
be governed by a single protocol.  Rather, different communities of actors that
belong to the same system, but engaged
in different types of activities, would be required to operate subject to
different protocols. And such communities often need to interact with each other.  Therefore, a PEM needs to support multi-community
systems, where different communities, operating under different protocols, need
to be able to interact,  subject to some constraints. Moreover, as we have shown \cite{min18-1},
effective modularity of  the set of protocols, that thus govern a complex
system, 
can be achieved by organizing them  into the so-called \emph{conformance
  hierarchies}.
                   
\item \emph{Dependability and Security}: By this we mean the degree to which a PEM can defend
itself against failures and programming errors (dependability) and  against attacks (security).
Dependability and security are, of course, critical for
many applications.
\end{enumerate}
\noindent
It is worth pointing out that the satisfaction of any of these qualities is not
 a zero/one predicate, so our objective is a \emph{substantial} satisfaction of
these qualities.

\section{On the State-of-the-Art of Protocol Enforcement Mechanisms}\label{art} In
the following two sub-sections we consider two existing protocol-enforcement
mechanisms mentioned in the introduction, evaluating them
in terms of the qualities described above, and finding both of them wanting.

\subsection{The Blockchain-based Smart-Contract Mechanism}
This type of  mechanisms---which was inspired  by a  1997  paper
\cite{sza97-1}  by Nick Szabo---became very popular recently, mostly for
financial and commercial applications (see \cite{pet16-1}).
The main characteristic of smart-contracts is that the enforcement of  contracts is carried out
over a \emph{blockchain}, which provides this mechanism with a high level of
security.

 But smart-contracts do not satisfy well the other three qualities listed above.
First, the latency of a smart-contract  cannot be shorter than the time in takes to reach consensus---a
fundamental element of blockchains.  And this latency is quite substantial--it
 is currently of the order of a few minutes under the various implementations of smart-contract, and it probably cannot be made much shorter than a few seconds. 
This means that smart-contracts \emph{cannot be used for many time-critical
applications}.

Second, the smart-contract mechanism is not scalable, as is frequently admitted
by many researchers and developers of such systems. And although many, like
\cite{vuk17-1}, are working on  reducing the level of unscalability, the lack of
scalability is inherent in the blockchain-based  smart-contract mechanisms. This is because,
despite the distributed nature of the consensus, regarding which block to admit
to the various copies of the blockchain, the enforcement itself is essentially centralized, and
linear, for the following reason:  A new block cannot be
admitted to the blockchains, without sacrificing security, before the previous block is resolved. Now suppose
that it takes $T$ seconds to select a block (via a distributed consensus
mechanism), to be admitted into the distributed blockchain, and to be resolved 
according to the contract at hand; and suppose that the average number of
transaction  that can be included in a block is $B$. Now, if more than $B$ new
transactions arrive, on average, to the blockchain in $T$ seconds, then the
length
of the queue
of transactions waiting to be processed will increase linearly in time.
And the latency will
increase, proportionally with the length of the queue. So, such a  mechanism  is
inherently unscalable.

Third, blockchain-based smart-contracts cannot handle really complex
systems, such as federated enterprises, supply chain, and health-care systems.
Such systems are composed of many different communities operating under
different interdependent contracts which often need to
inter-operate. But despite some recent attempts to make blockchains
interoperate, such as by the Cosmos project, none of them provides a practical and general 
solution to this problem. The Cosmos project, in particular, features a hub
blockchain involved in all interoperations. And it seems to us that the use of such a
hub would 
decrease  the scalability and the security of smart-contracts.

\subsection{The Law-Governed Interaction (LGI) Mechanism:} This mechanism, which is discussed in some detail in
\secRef{LGI}, satisfies most of the qualities we required in \secRef{quality}.
This includes very short latency, high level of scalability, and a very general
concept of controllable interoperability.
But although LGI is reasonably secure---arguably more secure than the
centralized access control mechanisms---LGI has a  serious security weakness
described in \secRef{heel}, which this paper aims to resolve.
We sometime
 refer to this weakness of the security of LGI as its \emph{Achilles heel}.

\section{An Outline of LGI, and  its  \emph{Achilles Heel}}\label{LGI}

This section is an outline of the LGI mechanism, which should suffice for the
understanding of the rest of  this paper. But for a deeper and more detailed description of
LGI, we propose two, somewhat dated, sources: (a) a Journal paper
\cite{min99-5}; and (b) the manual of LGI \cite{min05-8}. We confine ourselves
here to the treatment of a single community of actors that are supposed to
interact with each other subject to a common LGI-law
\EL---such a community is called an  \EL-community, and we occasionally refer
to it as $C$.

The enforcement of an LGI-law is strictly \emph{decentralized}, as follows:
Each member $a$ of $C$ interacts with other members of $C$
via a private surrogate (called a \emph{controller}) that enforces law \EL\ over the interactive activities
of $a$. Such an enforcement is done
 \emph{locally}, with no knowledge of, or dependency on,
 anything that happens simultaneously at the surrogates of other members of $C$.
 (This locality
is the consequence of the nature of laws, as we shall see  below).
So, the enforcement
of a law \EL\ over the interactive activities of  members of the \EL-community  is
done in a decentralized manner, and thus  in parallel,
by the various surrogates of the members of
$C$. This enforcement is  very efficient, and inherently scalable.

The rest of this section is organized as follows:
\secRef{cos} describes the service  that provides actors with the
controllers that can serve as their surrogates;
\secRef{creation} describes how a given actor can become a member of
the \EL-community, for a given law \EL;
\secRef{law} outlines the structure of LGI-laws;
\secRef{examples} introduces two simple examples of laws;
\secRef{beyond} points out that LGI can handle far more complex systems than
is implied by the single community discussed so far.
Finally, \secRef{heel}  discusses the   security vulnerability
of LGI, which is what this paper is intended to relieve.

\subsection{A Trustworthy Controller-Service (CoS)}\label{cos}
LGI require the availability of  a set of authentic generic controllers that
are trusted to  operate as surrogates of arbitrary actors, subject to any well
formed LGI-law selected by  them.  Such a generic controller is denoted by $T$, which suggests that it
needs to be trusted to enforce correctly any LGI-law loaded into it.
A controller operating under a given law \EL\ is called an \EL-controller, and
is denoted by  \Tla{L}{}.

 There are
several ways for supplying such generic controllers.
One of this is to create a \emph{controller service} (CoS) that maintains set
of authentic controllers, and leases them to its customers, presumably for a
fee.
The CoS can use various techniques for guarding against corruption of
controllers.
In particular it can use the TPM (Trusted Platform Module) technology, or some
more recent variant of it. This is particularly easy to do because all generic
controllers have identical codes, which would have a single and stable hash.

To be trustworthy,  a CoS must  be managed by
 a highly reputable organization, which can vouch for the authenticity of its
 controllers.
Moreover,  the CoS should  provide each of its controllers with a digital
certificate signed by the CoS itself.
 This certificate is  used by the controllers to
    authenticate themselves as  genuine
    LGI-controllers.
    The CoS is the trusted computing base (TCB) of LGI.

    \subsection{The Concept of  an  \EL-Agent, and  the Formation of an
  \EL-Community}\label{creation}

  \subsubsection{An \EL-agent and its formation}
  Any actor $a$ can attempt to operate under a given law \EL, thus joining the
  \EL-community.
  It can do
  so by performing the following two steps: (i) acquiring a generic controller
  $T$ from the CoS; and (ii) adopting this controller to be its surrogate,
  subject to  law \EL.

  If the adoption of a controller $T$ subject to law \EL\ is successful, then this
controller would be
 denoted by \Tla{L}{a}, which means that
 $T$ now operates as the surrogate of $a$, subject to law \EL.  Now,  the pair
  $ \langle a, T^{\mathcal{L}}_{a} \rangle,$ is called an \EL-\emph{agent},
 which is denoted by $\overline{a}$.
  The mediator \Tla{L}{a} is called
 the \emph{private controller} (or simply the controller) of agent $a$, and the
 actor $a$, whose internal structure is irrelevant to this model,  is said to
 \emph{animate} agent  $\overline{a}$.
 
But note that the attempt of $a$ to form an \EL-agent may fail, 
 because law \EL\ may refuse to allow $a$ to do so.  For example, law \EL\ may
 require a password, or a certificate, to be submitted by $a$ in its
  adoption command. This would prevent any actor that cannot authenticate
  itself in the manner required by law \EL\ from operating as a member of the
  \EL-community.

\begin{figure}
\leavevmode
\epsfysize=1.0 in
\epsfxsize=4.5 in
\centerline{\epsffile{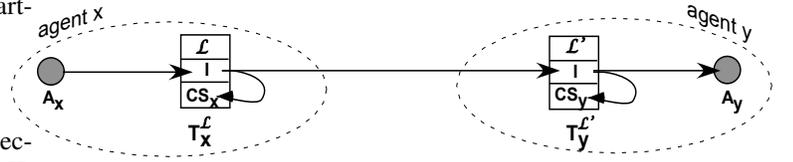}}
\caption{\emph{A pair of interacting agents, operating under possible different laws}
}
\label{fig-agent}
\end{figure}

\subsubsection{The Dual Mediation of Communication Under LGI}\label{ss-dual}  One of the significant aspects
 of LGI is that it involves dual mediation of every exchange of
messages between LGI-agents: one on the side of the sender of a message, and
one on the side of its receiver.  Specifically, the passage of a message from
an actor $A_x$ of an \EL-agent $x$ to an actor $A_y$ of an  \CAL{L'}-agent $y$,
 must be mediated first by the controller
 \Tla{L}{x}
 associated with  $A_x$, and then by the controller \Tla{L'}{y},
 associated
with $A_y$, as is illustrated in Figure~\ref{fig-agent}. 
 This is  a direct consequence of the locality of LGI-laws, which requires 
both the sender and receiver to individually comply with the law under which
each of them operates.  

  The dual mediation under LGI has several important
implications, not the least of which is that it facilitates interoperability by
providing flexible control over cross-interaction between agents operating
under different laws, as is further discussed in \secRef{beyond}.
Moreover, as has been shown in \cite{min99-5}, the dual control turns out to be
more efficient than centralized control, in many circumstances. 
A simple illustration of the nature of dual mediation, and some of its consequences,
is provided by the  example laws on \secRef{examples}.

  \subsubsection{On the  Formation of an \EL-Community}
  An \EL-community is created simply by the creation of the first \EL-agent by
  some actor $a$.  No special  initialization of an \EL-community is required.
And  the membership of such a   community evolves incrementally, by the creation or destruction of \EL-agents.
   The structure and dynamic behavior of such a community, which depends on the
   nature of its law,
   is explored in \cite{min00-3}.

\subsection{The Concept of Law Under LGI}\label{law}
 An LGI-\emph{law} \EL\ is formulated in terms of  three
 elements, defined with respect to a given \EL-controller $x$. The following is
 an incomplete description of these elements:

(1) A set $E$ of \emph{regulated events} (or, simply, \emph{events}) 
that may occur at  $x$.  $E$  includes, among others: (a) the \emph{adoption}
of this controller, under a given law \EL---which is the first event that occurs at $x$. (b) The arrival of a message at $x$, and
the sending of a message by it. Note that a message may arrive at $x$ mostly from three sources:
(i) from another controller; (ii) from the actor that adopted $x$; and (iii) as
    a result from some \emph{exception-condition}, which is reported to $x$ via a message.
    And (3) an event called
\emph{obligationDue} that has to do with the pro-active capability if LGI.
 This type of events  discussed  in \secRef{oblig}.

(2) The  \emph{state} $S_x$ of controller $x$,
 which  is distinct from the internal state of the actors that uses $x$ as its surrogate---of which the
 law is oblivious. The state $S_x$ is an unbounded (but often small) set of
 terms, whose structure is left unspecified here.

(3) A set $O$ of
 \emph{operations} that can
 be mandated by a law, to be carried out by the controller $x$  upon the occurrence
 of a regulated events at it. The set $O$ includes, among others: (a) replacing  the state  $S_x$ of
 $x$ with another state (or, if you will, changing the state $S_x$);
 and (b) sending some  messages, to anybody on the Internet. (Note that the sending of a message by
 $x$, as mandated by the law, would constitute a new event at $x$.)

Now, the role of a law  is to decide what should be done in response to
the occurrence of any regulated event at a controller  operating under it.
This decision, with respect to controller $x$,
is formally defined by the following mapping:
\begin{equation}
 E \times S_x \rightarrow  (O)^*  \times   S_x 
\label{eq-law-2}
\end{equation}

Using a  less formal notation: the law, when applies to a given controller $x$, is a function

\begin{equation}
law: (e,s) \rightarrow (m,ns)
\label{eq-law-3}
\end{equation}
\noindent
that maps any  a given pair $(e,s)$, into a \emph{ruling} $(m,ns)$.
Here  $e$
is an event that occurred at controller $x$, and $s$ is the state of $x$ at the time
of occurrence of this event. 
And the ruling that $x$ is to carry out, consists of: (a) a (possibly
empty) list $m$ of \emph{operations} that $x$ must execute (which are, most
often, messages to be sent);  and (b) a new
state $ns$ that   is to replace the current state of $x$.

This definition makes it clear that the law is strictly \emph{local}. Indeed, the event
that occurs at the controller and its state are local. And the ruling is to be
carried out locally. Of course, if the ruling calls for some message to be sent
to another controller, then this message would eventually have a non-local
effect. But the decision to send such a message is local.

 It should be pointing out that while  Formula~\ref{eq-law-2} is a 
definition of the semantics of laws\footnote{Modulo the fact  that  the sets $E$ of events and $O$ of
operations have not been fully spelled out here.}, it does not
 specify a language for
writing them.
 In fact, the current implementation of LGI supports  several
\emph{law-languages}, one of which is JavaScript.
 The choice of a law-language has no effect on the
semantics of LGI, as long as the chosen language is sufficiently powerful to
specify all possible mappings defined by Formula~\ref{eq-law-2}.

\subsubsection{Additional Observations about LGI-Laws and their Enforcement}\label{more-observations}

\paragraph{Non-Deterministic Laws} A law may be \emph{non-deterministic} in
that its rulings may include random numbers. Such laws are useful, in
particular, for the protocols involving randomness and tie-breaking. The
implementation of non-determinism under Cop is discussed in \cite{ccong}.

\paragraph{On the Interplay Between the Fixed Law and the Changing State of a
  Controller $x$}
On one hand, the ruling of the law may depend
on the current state of $x$,  on the other hand the
evolution of the state is regulated by the law---although it is driven by the
various event that occur at $x$, most of them coming from other controllers in a
given community.

\paragraph{About the Enforcement of Laws}
A controller $x$ deals with events occurring at it sequentially, and if several
events occur at $x$ at the same time, they will be handles in an arbitrary order.
Also, the ruling of an event $e$ is carried out atomically, before handling any
subsequent  event.

 \subsection{Two Examples of Laws}\label{examples}
 We introduce here two very simple examples of LGI-laws, called
 \emph{money transfer law} (\CAL{MT}) and  \emph{monitoring law} (\CAL{MO}).
 The formar law is stated
 formally in one of the law-languages currently supported by LGI, 
and the latter law is described informally.
We will return to these laws in \secRef{heel}, and also in \secRef{Cop}.

\textbf{A Money Transfer  Law (\CAL{MT}):}
This law provides an initial budget of \$1000 to every \CAL{MT}-agent (this is
done upon the adoption of a controller with  law \CAL{MT}).
And then, this law enables every \CAL{MT}-agent  to transfer to others any
amount of money smaller than or equal to its current budget.
A formal statement of this law---written in  the the law-language based on
CoffeeScript (a semantically equivalent variant of JavaScript)---is spelled out in     Box \ref{law:MT}. This law has three rules, each of them contains comments (lines
starting with \#) that explain its effect.

\textbf{A Monitoring (\CAL{MO}) Law:}
The following informally stated law,  called  \CAL{MO}, establishes a systematic monitoring scheme
for all communication within a given community.
\begin{enumerate}
\em
\item  When any new \CAL{MO}-agent is created---by an actor adopting a
       controller after inserting law \CAL{MO} into it---the controller of the newborn agent
       would send a
	message to the designated monitor, essentially recording its own  birth.
\item  Whenever an \CAL{MO}-message is
	sent, a copy of it, along with the addresses of the sender and   its target, is sent to
	the monitor.
\end{enumerate}

\begin{lawscript}{Money Transfer Law}{law:MT}\label{MT}
Name: MT
LawScript: CoffeeScript

# (R1) When the controller is adopted,
#      initialize the agent's budget to 1000.
UPON "adopted", ->
  DO "set", key: "budget", value: 1000
  return true

# (R2) An agent can send any positive amount of money
#      to another agent provided that the amount
#      is not greater than its budget,
#      then the amount will be deducted from its budget.
UPON "sent", ->
  if @message > 0 and @message <= CS("budget")
    DO "set", key: "budget", value: CS("budget") - @message
    DO "forward"
    return true

# (R3) When an agent receives a positive amount of money,
#      the amount will be deposited to its budget.
UPON "arrived", ->
  if @message > 0
    DO "set", key: "budget", value: CS("budget") + @message
    DO "deliver"
    return true
\end{lawscript}

\subsection{Beyond  Singleton Communities}\label{beyond}
So far we have discussed the case of an singleton \EL-community, whose members
interact with each other subject to a common law. But LGI is
far more general than that, in the following ways, in particular:
First, LGI can handle any number of  communities, operating under different laws.
Second,  LGI  can enable members of different such communities, say $C1$ and
$C2$---each operating under its own
law---to interact with each other in a regulated manner. This can be done by having
the laws of each of these community specify the condition under which its
members can interact with each other.
And third, LGI enables the organization of  a
set of laws that   collectively governs a
single system, into  a coherent ensemble called a \emph{conformance hierarchy} $H$.
$H$ is  a
tree of laws  rooted by a law called \law{R}. And 
  every law in $H$, except of  \law{R} itself,  conforms transitively to its
 superior law. Moreover the conformance
 relation between laws in $H$ is inherent in the manner in which $H$ is constructed, requiring no extra validation.
For a formal definition of such an hierarchy of laws see \cite{min03-6}, and
for a recent application of it to complex systems see \cite{min18-1}.

\subsection{The Security Vulnerability of LGI}\label{heel}
  Even if the CoS of LGI does its utmost to maintain and protect authentic
    controllers  there is, of course, no way to ensure that controllers cannot
    be corrupted and would violate the the law under which they are supposed to
    operate. Such a corruption may be the result of an attack on a controller,
    either by an insider of the CoS, or by an outsider who discovered some
    vulnerability in the code of controllers.  We are concerned here mostly
    about the resulting \emph{Byzantine behavior} of controllers, and not about
    their fail-stop type of failure, which can be handled effectively by LGI

 The possible failure of individual controllers may be  considered an
acceptable risk in distributed computing, as it poses a smaller risk than that of the
corruption of a central reference monitor commonly used in access
control. Indeed, the corruption of a central reference monitor can  endanger
an entire system, while the corruption of  a few controllers usually have  a
more local effect.

Yet, in some cases
 a Byzantine failure of even a single
 controller may cause a serious damage to the community at large.
 A case in point is the money-transfer law
    presented in \secRef{examples}. If a single controller has been corrupted,
    it may be able to distribute a large amount of fake money among other
    members of the \CAL{MT}-community, without raising any suspicion, at least
    for a while.     The  \emph{Achilles heel} of LGI is that it provide no
    general means for even detecting corrupt controllers.

Our approach for resolving this   \emph{Achilles heel}, thus
protecting a system from the misbehaviors of its
controllers, is the following:
We  provide a general mechanism that detects
quickly and reliably any failed
controller, right after it first failed to satisfy the law under which it
operates; and then to recover from such a failure. 
This mechanism, called \emph{detection \& recovery} (or $D\&R$), is 
 the subject of the rest of this paper.

But first, we should make the following observation:     
        There is, of course, a  very general  technique for handling Byzantine
    failures, see  \cite{cas00-1} for example. In principle, this techniques
    can be applied to every  controllers of LGI.
    But this would be prohibitively too inefficient and 
    expensive for most potential applications of LGI.

 \section{The LGI-Based Cop Mechanism}\label{Cop}
 Cop  carries out two complementary
 functions in enforcing a given law \EL\ over an \EL-community $C$.  One function
 is the enforcement, \emph{per se}, of law \EL\ over the interactive activities
 of the members of $C$. The other function is the speedy detection of
any failure of  an \EL-controller to enforce law \EL,  followed immediately by
the recovery from this failure, which includes the
 repair of the failed  controller, and by the resumption of its operation.

These dual functions are carried out by two disjoint processes that operate in
concert.  One process is the enforcement of law \EL\ over the \EL-community, which
is done by means of the LGI mechanism,  outlined in
\secRef{LGI}\footnote{There is just one difference between the LGI version used
in Cop, and the older LGI---it is the implementation of the concept of \emph{enforced
  obligation}},
which provides LGI with an important proactive capability (cf. \cite{ccong}).
The other process, which operates off-line of the enforcement mechanism, is the detection of any misbehaving (or failed) controllers, and
the recovery from such failures. This is done by a mechanism called
D\&R, for  “Detection and Recovery”.
  The D\&R part of Cop is  the main subject of the rest of this paper.

 It should be noted that---for simplicity---most of our description of Cop
 involves the treatment of a single  isolated  \EL-community, whose members interact only with each
 other subject to law \EL. But as we shall see, Cop can handle any number of such
 communities. Moreover, as explained in \secRef{interop}, Cop is not limited
to dealing with isolated communities. Rather,
  like LGI itself, Cop can
govern complex systems constituted of many interacting communities, 
inter-operating with each other subject to a conformance-hierarchy of laws.

We conclude this section with a description of the main
components of Cop, and of the roles they play.
The operations of the  D\&R mechanism, is discussed in
\secRef{detection}.

 \subsection{The Components of the Cop Mechanism}\label{structure}
 The Cop mechanism is composed of three types of components that operate in
concert. 

\begin{enumerate}
\item  The \emph{controller provider} (CP), which is a variant of the CoS of
 LGI. Besides the maintenance of generic controllers, as does the CoS, the CP participates
actively in the operations of the D\&R mechanism.
\item The \emph{ledger} $D_L$ that maintains a record of the interactive
      behavior of all \EL-controllers---i.e., the controllers that serve the \EL-community.
\item The \emph{inspector} $I_L$, which performs the inspection of the
      interactive behavior
of all \EL-controllers in order to detect any  failure of any one of
them, and to initiate the recovery of such failures.
\end{enumerate}
\noindent
Note that  while the  Cop mechanism has just one CP, which can handle any
number of communities,  each active \EL-community is served by its own pair ($D_L$, $I_L$) of
ledger and inspector, respectively. 
 Note also, that the various ledgers and
inspectors are to operate  on different hosts, then the hosts used by CP.
These three types of components are  described in details in the following three subsections.

\subsubsection{\textbf{The Controller Provider (CP)}}\label{CP}
The CP plays two kinds of roles.
First, like the controller-service  (CoS) of LGI, CP maintains generic
controllers, and provides them to its clients. In that, CP is a more reliable
version of the CoS, as argued below.
Second, CP plays an important role in the operation of the D\&R mechanism.
These two  roles are discussed in the following two paragraphs.

\paragraph{The Maintenance of Generic Controllers} 
The main structural difference between the CP and the CoS, in this respect, is that under the
CP,  controllers are encapsulated in Linux \emph{containers} \cite{mer14-1},
hosted by a group of servers we call \emph{CPnodes}, which are managed by the CP.

 The CP can maintain any number of CPnodes, and each CPnode can host several
 hundreds of generic controllers, depending on the applications using them.
 And the controllers residing on any given
 CPnode  may end up operating under different
 laws, and thus serving different communities.
All the controllers resident in a given CPnode are 
 managed by a \emph{local
manager} running on this CPnode---some of the functions of the local manager
will be discussed in due course.  The CP as a whole is managed by a
\emph{global manager} running on a distinguished CPnode of the CP. (And it would
be useful, and quite elegant although not entirely necessary, for all these
managers to interact with each other, subject to suitable \emph{management-law}
under LGI---this is not done in present prototype of Cop.)

The encapsulation of controllers in containers has several 
advantages. In particular, this architecture enables the imposition of limits
on the use of various resources---such as CPU, memory and communication---by
 individual  controllers. Such limits  can be imposed by the local manager
 of the CPnode and  dynamically adjusted by it. The ability to impose such  limits
would make   controllers
  more robust because it can help
to prevent an individual  controller from hogging resources, thus
preventing others from operating effectively, or at all.

\paragraph{The Role that CP Plays in the Operations of the D\&R Mechanism} 
 The CP carries out two functions  that are essential to the D\&R mechanism, as
 follows:

\emph{(b.1) Supplying the ledger  $D_L$ with the information it needs:}
As we shall see later, the detection of the failure of
controllers  by D\&R requires that all the 
 events that occur  at every
 \EL-controller, and all the operations  carried out by it,
 be logged correctly on the ledger $D_L$ (cf. \secRef{ledger}).
  Unfortunately, we cannot trust the controllers themselves to log their own
  events and operations, because anyone  of them may be corrupted.
 So, such logging is to be done by the local manager of every CPnode, by intercepting  all the  messages sent or received by every
      controllers that resides on the CPnode in question.
      And note that the sending or
receiving of a message by a given controller corresponds to some event that occurred in
it, and/or some operation carried out by it.
 The events and operations thus logged in ledgers are  time-stamped, using the
local time of the CPnode on which the controller resides;  and they  identify the controller in
question.

Finally, it is important to point out that
 controllers operating subject to a given law \EL, thus serving actors
belonging to the \EL-community, may reside on different CPnodes. The managers of these
CPnodes would  log events that occur of \EL-controllers on the $D_L$ ledger.
Conversely, the manager of a given CPnode, which may host controllers operating
under different laws, would have to log events on different ledgers.

\emph{(b.2) Repairing failed controllers:}
As discussed in \secRef{recon},  the reconstruction of a failed \EL-controller  is carried out by the CP---following
 an instruction by the inspector  $I_L$.

\subsubsection{\textbf{The Ledger}}\label{ledger}
As has already been pointed out, there is one ledger $D_L$  per an
\EL-community, which
is designed  to maintain entries representing two kind of items: (a) the
\emph{events} 
that occurred at any given 
\EL-controller $z$, which represent messages obtained by $z$ from various sources;
and (b)  the
\emph{operations} carried out by $z$, most of which  represent messages sent
by it to various targets.
These entries are supplied to the ledger by the CP, as pointed out above.

Such a ledger can have various architectures, providing different level of security and
efficiency to Cop. It can, in particular, be some form of blockchain, 
 such as under Ethereum \cite{pet16-1} or  under HyperLedger \cite{vuk17-1}.
 Or it  can  be just a file, perhaps replicated, which a given
organization maintains as part of its TCB.   We will not discuss here the pros and cons of the  various   implementations
of the ledger.

\subsubsection{\textbf{The Inspector}}\label{inspect}
The objective of an  inspector $I_L$, associated with a given \EL-community,
is twofold: (a) to inspect the interactive behavior of  all the \EL-controllers, in
order to detect any failure of any of them; and 
(b) to initiate  the recovery  measures for failed controllers.

 It is clear that the detection  of a corrupt controller, and its recovery, needs
 to be done as quickly as possible---because the longer a corrupt controller is allowed
 to operate, the more damage it can do---damage
 that may be hard to reverse.
 The manner in which the inspector operates is
 discussed in \secRef{detection}.

\begin{figure}
\leavevmode
\epsfysize=4 in
\epsfxsize=3.7 in
\centerline{\epsffile{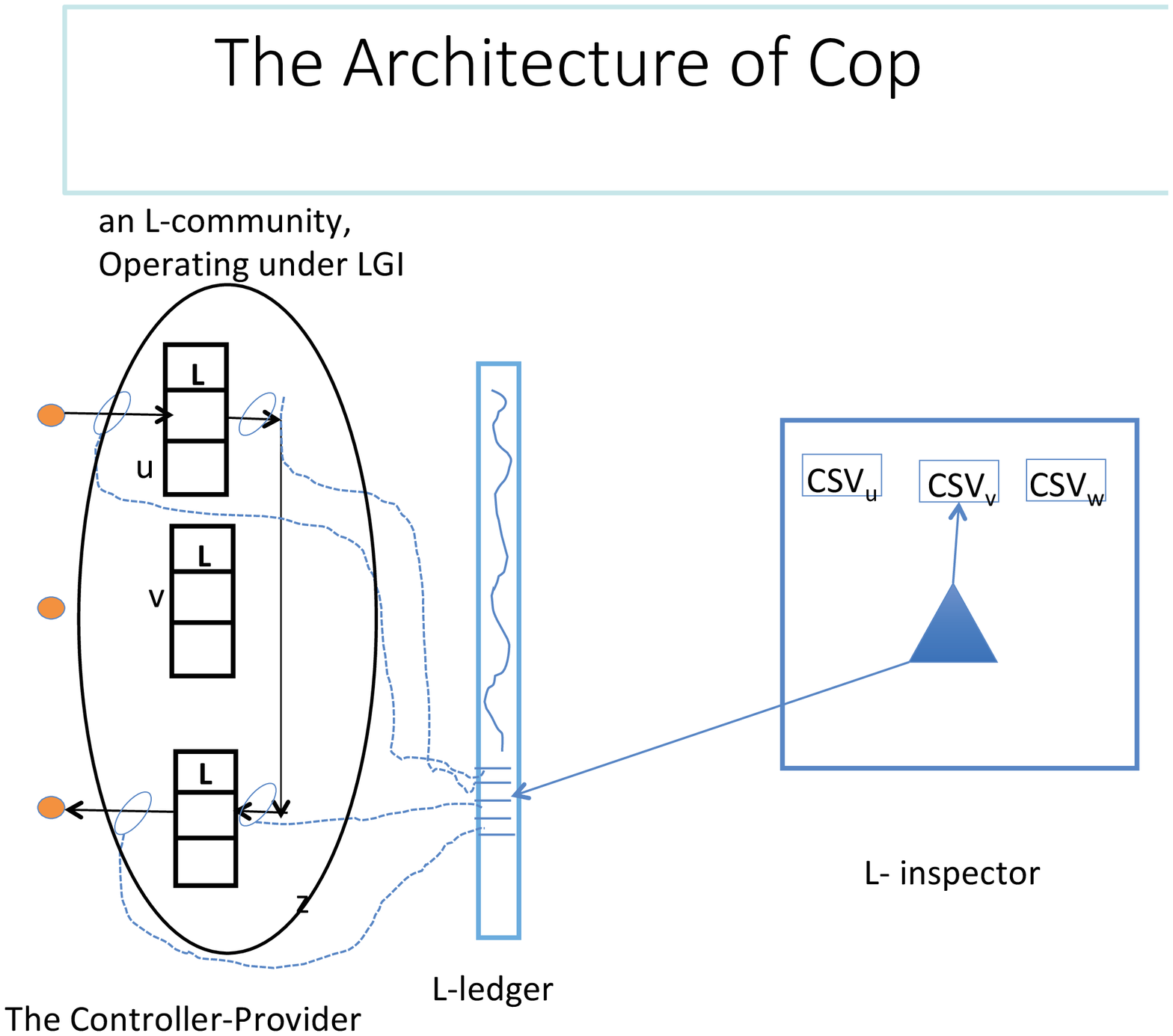}}
\caption{A Schematic Depiction of the Operation of CoP
\label{fig-cop}}
\end{figure}

\subsubsection{The Architecture of Cop}\label{arch}
Figure~\ref{fig-cop} provides a schematic depiction of the overall architecture and
behavior of Cop. This figure depicts the treatment under Cop of the interaction
 between two actors  operating under a law denoted by $L$, via a pair of
controllers residing in a given controller-provider (CP). The various events
and operation generated by this interaction are intercepted by the the CP and
sent to the L-ledger, which is inspected by the L-inspector.

\section{Inspection: The Process of  Detection and Recovery   of Failed
  Controllers} \label{detection}
This section starts with some introductory observations about the inspection
process.
 Then,
\secRef{single} discusses the inspection of a single \EL-controller,
\secRef{scaling} discusses the inspection of all the controllers serving a
given  \EL-community,
and \secRef{recovery} focuses on the recovery from the  failures of controllers.

\subsection{Introductory Observations}\label{observations}

\subsubsection{Locality: the Key to Effective Detection  of Failed
  Controllers}\label{locality}
The detection of the failure of controllers, by inspecting their interactive
behavior,  may seem to be a daunting and time consuming process,  because the 
 behavior of each controller depends on its interaction with others.
 This seems to suggest the need for global analysis of the process of
interaction between all controllers for detecting any violation of the law by
any of them.

 Fortunately, no such global
analysis is required, due to the inherently local nature of the laws of  LGI (cf.
 \secRef{LGI}). This aspect of laws under LGI is what
enables it to enforce its laws locally, and thus efficiently and scalably.
And since the enforcement is local,  it follows that
	non-compliance with the law can also be dealt with locally---at each
	controller, independently of all other controllers.   This
	simplifies the process of inspection enormously, and makes it very
	efficient and scalable. Thus, the locality of laws is the key for
 effective detection of the failure of controllers

\subsubsection{An Invariant of the Inspection Process} The inspector $I_L$ maintains a 
 variable  called  $CSV_x$---which stands for \emph{Controller State-Variable}
 of  $x$---for every \EL-controller
       $x$ being inspected.  $CSV_x$ is computed, by the inspector, such that
 the following invariant is maintained:
 \begin{quote}
   The value of  $CSV_x$ is
       the \emph{correct} state of the controller $x$ associated with the current event
       that occurs at $x$.
 \end{quote}
 \noindent
By ``correct state of the controller'' we mean the state that a good
controller $x$ would have
associated with the event that occurs at it. Of course,  a  failed, Byzantine,
controller may have an arbitrary 
state, whose value cannot be predicted. 
       We will see later how this invariant is maintained by the  inspector.

 \subsubsection{The Initialization  of an \EL-community Under Cop}\label{init}
We described  in  \secRef{LGI} the process of the creation and incremental
development of an \EL-community under LGI, which require no formal initialization.
 But under Cop, this process needs to be prefaced with the following
 initialization steps: (a) the
formation of law \EL; (b) the creation of an empty ledger $D_L$; and (c) the creation
 of  inspector $I_L$, which is given the law under which it is to operate.  Note that initially this community has no members, and
 no activity to be recorded on the ledger $D_L$. 
 The inspector $I_L$ starts examining the ledger, but doing nothing else until
 it finds some entries on the ledger to be inspected.

\subsection{The Inspection of a single \EL-Controller}\label{single}

Broadly speaking, the process of inspection of a controller $x$ by the
inspector $I_L$ starts when $x$ becomes an  \EL-controller. This happens when
$x$ is adopted by some actor $a$, to operate subject to a law \EL.
 This action by actor  $a$ triggers the so-called  \emph{adopted}
event at $x$, which is the first event in its lifetime.
When this event is intercepted by the CP and stored in the ledger
 $D_L$, it is observed by the inspector $I_L$ and inspected, as  described in
  \secRef{inspection-1} below. After this initial inspection of $x$, its inspection
 continues recursively  in response to   consecutive  events
 at $x$  until $x$ quits---as described in \secRef{inspection-2}.

For the detailed discussion of this process it would be helpful to recall the description of an LGI law
provided by Formula~\ref{eq-law-3}, namely: a law is a function
\begin{equation}
law: (e,s) \rightarrow (m,ns)
\end{equation}
\noindent
that maps any  a given pair $(e,s)$, into a \emph{ruling} $(m,ns)$.
Here  $e$ is an event that occurred in controller $x$, $s$ is the state of $x$ at the time
of occurrence of this event. 
And the ruling that $x$ is to carry out consists of: (a) a (possibly
empty) list $m$ of \emph{operations} that $x$ is to perform (which are, most
often, messages to be sent);  and (b) a new
state $ns$ that   is to replace the current state $s$ of $x$.

\subsubsection{The initial Inspection of $x$}\label{inspection-1}
This inspection  starts when the  inspector $I_L$---which continuously scans
the ledger $D_L$ for new
entries---notices  the \emph{adopted} event at $x$ on the ledger, which prompts
it to carry out the following sequence of steps.

(i) The inspector  creates the variable  $CSV_x$ of $x$, whose value, in general, is
      to be 
the  correct state of the controller $x$ associated with any given event
       that occurs at $x$. The initial value of 
        $CSV_x$ is set to be an empty
      set, because, as stated in \secRef{creation}, this is the state of every
      newly formed controller, before its \emph{adopted} event is evaluated.
      
(ii)  The inspector computes the ruling of law \EL\ for the pair   $(e,s)$,
       where the event $e$ is the adopted event and the state $s$ is the
       current value
       of  $CSV_x$, which is the empty state, as pointed up above. The ruling $(m,ns)$  
       defines what a good $x$ is expected
       to carry out, atomically.

(iii) Getting this ruling, the inspector changes the value  of  $CSV_x$ to
      $ns$, which is clearly the correct state of $x$ at this point, because it
      is mandated by the law.

(iv)   The inspector will  now verify if the controller $x$ carried out the required list of
     operations in $m$---no more then in $m$ and no less than in it. To do this, the inspector needs to compare $m$ to the list $m'$ of operations
     found on the ledger between the adopted event of $x$ and its next
     event. (This must be the place for these operations, because the ruling of
     the law must be carried out atomically, and thus before the  occurrence of next event at
     $x$.)
     Now, the inspector has two possibility to consider:

First, if $m'$ equals to $m$, then the inspector
     concludes that  $x$ operates correctly.

 Second, if $m'$ differs from $m$, 
 the inspector concludes that $x$ just failed, and it will commence the
appropriate recovery procedure, as described in \secRef{recovery}.

\subsubsection{The Recursive Inspection of $x$}\label{inspection-2}
Suppose that inspector $I_L$ inspected a sequence of events  $(e_1, e_2, ..., e_n)$ that
occurred at $x$, without detecting any failure---where $e_1$ is the very first
event that occurred at $x$, namely the \emph{adopted} event discussed above.
The inspection will continue, recursively, to the next event, if any, as  follows:

(i) $I_L$  will compute the ruling of the law for  the pair $(e_{n+1},s_n)$, where $e_{n+1}$
is the new event at $x$ found on the ledger, and $s_n$ is
the current value of  $CSV_x$ which is obtained during the previous step
of this recursion.
Suppose now that the ruling of  the law for  the pair $(e_{n+1},s_n)$ is the pair $(m_{n+1},ns_{n+1})$,
whose structure was described before.

(ii) Getting this ruling, the inspector plants $ns_{n+1}$ as the new  value of 
 $CSV_x$.

(iii) The inspector will  now  check if $x$ carried out the required list of
     operations $m_{n+1}$. To do this it needs compare this list to the list $m'$ of operations
     found on the ledger immediately after  the $e_{n+1}$ event of $x$, and before the
     next event  that occurred at $x$, if any.
     Now, the inspector has, again, two possibility to consider,
 in a direct analogy to the two
possibilities  it had in the very first inspection of $x$:
First, if $m'$ is equals to $m_{n+1}$, then  the inspector concludes
that  $x$ operates correctly.
And second, if $m'$ differs from $m_{n+1}$, 
then the inspector concludes
 that $x$ just failed, and it will commence the
 recovery procedure from this failure, as described in \secRef{recovery}.

It is worth pointing out, again, that the correctness of the value of $CSV_x$ as the true
state of the controller which should be used for the evaluation of the law,
is an \emph{invariant} of this recursive inspection.

\subsection{The Scalable Process of Inspecting of all \EL-Controllers}\label{scaling}
A single inspector can, in principle, inspect all the 
the \EL-controllers that serve a given \EL-community.
Such an inspector would maintain the CSV of all active \EL-controllers, in
a single address space. And it would  inspect all these controller virtually in parallel.
But there is a potential problem  with this modus operandi of inspection. Namely, when  the size of the community
increases, the   latency between the failure of a controller
and the detection of this failure grows, roughly linearly.  In other words, such
inspection is unscalable with respect to this latency. 
And  the longer the latency is, the more opportunity a failed controller would have to send
illegal messages that can cause serious damages to the system, and which may be
very hard to recover from.

Fortunately, it is very easy to make the process of inspection scalable, as follows:
If the set of 
\EL-controllers in a given community is considered too large, in the sense that it
produces overly large latency,  one can divide  this set to any number
of subgroups that can be
handled by different, but identical, copies of  inspector that operate in
parallel.
Such division can be done dynamically, while the inspector operates.

\subsection{Recovery from the  Failures of Controllers}\label{recovery}
We  consider here two  complementary kinds  of recoveries.
The first is the resumption of the proper operation of a  failed controller by
its reconstruction.
The second deals with the affect of the failed controller  on other
  parts of the system.

  \subsubsection{The Reconstruction of a  Failed \EL-Controller}\label{recon}
The  reconstruction of a failed controller
 $x$ is prompted  by the inspector, and carried
  out by the local manager of the CPnode that host $x$.
  The reconstruction is carried out by the following  sequence of steps:

(i)  Controller $x$ is replaced---without changing its address---with an
authentic generic-controller provided by the CP.
(ii) Law \EL\ is planted into $x$, making it into an \EL-controller.
(iii) The latest value  of $CSV_x$ is planted into $x$.
And, (iv) the reconstructed controller $x$ is reactivated.

\subsubsection{Regarding the Affect of a Failed Controller  on Other
  Parts of the System}

Note that an \EL-controller is recognized by the inspector as failing,  after
 it operated illegally, relative to   law \EL. Such an illegal
 operation needs to be corrected.
Now, the ruling of a law is a list of zero or more  operations that are to be
carried out by the controller. The failure to carry out a given ruling consists
of one or more  instances of two types of illegality: (a) an
\emph{illegal inaction},
namely, the failure to carry out one of the operations in the
ruling of law \EL; and (b) an \emph{illegal action}, namely, carrying out an operation
not in the ruling of  law \EL. These two types of illegalities require
different handling, discussed below:

\paragraph{The Handling of Illegal Inaction}
What one needs to do to recover from this kind of illegality is to carry out the operation
required by the ruling. The Inspector does that right after discovering a 
failing controller, for every one of its illegal inactions, if any.

\paragraph{The Handling of Illegal Action}
An  illegal action means that the failing controller $x$ sent some message to
some controller $y$---an operation not mandated by the  law. Such a message
cannot be stopped, and its affect on $y$, and possibly on other members of the
\EL-community can be quite complex and possible serious---as exemplified by the
discussion of law \CAL{MD} in \secRef{heel}. The recovery from  such a
an illegal action may depend on the nature of the message sent, on the law, and
 on the application at hand, and is  not a simple matter.

 But Cop can help  in recovering from such an illegal action, by having the inspector  
 notify some designated
manager---one associated with the \EL-community---which can analyze the
situation and decide how to rectify the problem. The operation of such a
manager  is beyond the scope of this paper. But it is worth pointing out that
such a  manager may need to examine parts of the $D_L$ ledger, to determine the
extent of the affect of the illegal message.

\section{Additional Aspects of Cop}\label{aspects}

\subsection{The Controllable Interoperability Under Cop}\label{interop}
	As explained in \secRef{beyond}, the LGI part of Cop---which is the part that
	enforces laws---provides for a sophisticated kind of interoperability.
        This capability of LGI extends to D\&R, and thus to  Cop.
        
Consequently, Cop can be used to govern the interaction between the the members
of disjoint communities. Moreover, can be used for governing  very complex
systems, such
as federated enterprises, which
	comprises of a collection of communities operating under different laws
	that inter-operate with each other. Furthermore, it is possible to
	organize such inter-operating laws in a \emph{conformance hierarchy}, which provides a way
	to control which communities can inter-operate with each other, and how.
        This provides Cop with considerable generality.

\subsection{The LGI's Concept of Enforced Obligation, and its Treatment under Cop}\label{oblig}
The  concept of  \emph{enforced obligation} (or ``obligation'' for short)  provides LGI with an important
\emph{pro-active capabilities},  invaluable for security and for fault
tolerance.
 This concept can be used, for example, to
 ensure that resources will not stay locked indefinitely, or to penalize book
 borrowers that do not return a book in the appointed time.

Informally speaking, an obligation \emph{incurred} by a given controller, serves as a kind
of \emph{motive force}, which ensures that a certain action (called
\emph{sanction}) is carried out by this controller, at a specified time in the
future (the deadline), when the obligation is said to
\emph{come due}---provided that certain conditions on the state
of the controller are satisfied at that time. This mechanism is governed by the law
in question.

 Specifically, a controller $x$ incurs an obligation by the execution, as part
 of the ruling of some event, of an operation \textbf{imposeObligation(oName,dt)},
where  \TT{oName} is the name of the obligation
and \TT{dt} is the time period after which
the obligation is to come due.
When this obligation comes due, after \TT{dt} seconds, the event
 \textbf{obligationDue(oName)} would occur at controller $x$. The
 occurrence of this event would cause the controller to evaluate the ruling of
 the law for this event, and to carry out its ruling.  The ruling of
 the law about an \TT{obligationDue(oName)} event is, thus, the \emph{sanction}
 for obligation \TT{oName}.

 But this concept of LGI, as is, cannot be supported under Cop. Because it
 relies on the controller itself to determine when the
 \TT{obligationDue(oName)} event occurs. And our D\&R mechanism does not, and
 cannot, rely on the controllers themselves.
 So, we implement the concept of obligation under Cop by making two changes
 to the obligation mechanism of LGI, without changing its semantics.
 First, the operation \textbf{imposeObligation(oName,dt)} causes a message with
 this text to be sent to an \emph{obligation-server} implemented in the CPnode  
 where the control $x$ in question resides. When the deadline for this
 obligation arrives, the obligation-server is programmed
 to send the message  \textbf{obligationDue(oName)} to $x$.

 Second, the  \textbf{obligationDue(oName)} message
 sent by the obligation-server,  is viewed by  the receiving controller as an
 event that  would be handled  just as the internal  event \textbf{obligationDue(oName)}
 is handled by the original LGI. That is, 
the controller would evaluate the ruling of
 the law for this event, and then carry out its ruling---thus preserving the
 original semantics of the enforced obligation under LGI.

 It should be pointed out that this is the only change in LGI we make in order
 to incorporate it in Cop.

\subsection{An Implementation of a Prototype of Cop, and its Testing}\label{testing}
We have implemented a fully functional prototype of Cop, and tested it.
The following is a summary of the results of this testing of 
 the correctness and the performance of the  D\&R
mechanism of Cop. (The correctness and performance of the LGI part of Cop has
been tested many times in the past, and these measures did not change under Cop.)

To test the correctness of  D\&R---which is, in a sense, the test of the
security of Cop---we've
    applied this prototype to  multiple types of test cases, with
    different laws.
    We found
    \emph{no false negatives}, i.e., all failed controllers were discovered.
    And  we found
\emph{no false positive}, i.e.,  no correct controller were reported as a failed.
Moreover,  after the recovery of failed controllers, they all  behaved correctly.

We have also measured the performance of the prototype of the D\&R
mechanism. We found (a) the latency
of discovering a failed controller to be 6 seconds, on the average; and (b) the
latency of the recovery of failed controllers to be 2 seconds, on the average.

But these performance measures have a limited value.
First, because it is just a prototype which we tested, whose code is
not optimized, in particular, it is written in  Python 3---a scripting
language with relatively low performance. And second, because our
experiments were done on a relatively weak hardware, which did
not allow us to experiment with large communities.
We expect that our latency result would be  reduced by an order of magnitude,
with optimized code, run on a stronger hardware.

\subsection{On the Trustworthiness   and Security of the D\&R   Mechanism}\label{security}
The D\&R mechanism has been designed for protecting against the possible failure of
controllers---which we call the  Achilles
heel of LGI.  But this protection can be effective only if the  D\&R
  Mechanism itself---consisting of the CP, the ledgers, and the Inspectors---is
  secure. That is, these three components must constitute the trusted computing
  base (TCB) of D\&R.  We believe that our current design of these components
  of D\&R of makes these components reasonably secure. And their security can
  be enhanced via various traditional means, including TPM and related
  technologies.
Such security enhancements are beyond the scope of this work.

\section{Conclusion}\label{conclusion}
  This paper introduces a scalable and secure protocol-enforcement mechanism,
  called Cop, which fulfills important qualities for such mechanisms, including  low latency, high
  scalability,  general interoperability and security. It is thus applicable to
  a wide range of applications, including  small
  and large systems,  time-critical systems, and
systems-of-systems.

 Cop enforces protocols via the existing Law-Governed
Interaction (LGI) mechanism,
coupled with a new protective layer called D\&R that discovers
any failed LGI-controller and repairs it---which is done soon after the failure occurs.
The  D\&R layer of Cop operates off-line relative to the enforcement by LGI,
and it significantly enhances the
dependability and security of the enforcement.

  We have implemented a fully functional prototype of Cop, and verified
  experimentally  its 
  correctness. But the evaluation of the performance and security of Cop, particularly when
  it is applied to large scale systems,
would require an optimized implementation of Cop, and
a sufficiently powerful hardware for it to run on.

  \appendix[Substantiating the Main Claims of this Paper]\label{append}
We have created a fully functional prototype of Cop, and tested it
(cf. \secRef{testing}).
This prototype will be made available on the authors' websites.


\bibliographystyle{plain}
\bibliography{../../writing-tools/biblio}
\end{document}